\documentstyle[12pt]{article}
\newcommand\be{\begin{eqnarray}}
\newcommand\ee{\end{eqnarray}}
\begin{document}

\begin{tabbing}
\`SUNY-NTG-92-15\\
\`July 1992
\end{tabbing}
\vbox to  0.8in{}
\centerline{\Large \bf Photon production through $A_1$ resonance }
\centerline{\Large \bf in high energy heavy ion collisions}
\vskip 2.5cm
\centerline{\large L. Xiong, E. Shuryak  and G. E. Brown }
\vskip .3cm
\centerline{Department of Physics}
\centerline{State University of New York at
Stony Brook}
\centerline{Stony Brook, New York 11794}
\vskip 0.35in
\centerline{\bf Abstract}
\indent
  Electromagnetic radiation from excited hadronic matter
is one of the best ways to study the  properties of the matter.
 Considering various
processes in the hadronic gas, Kapusta et al \cite{Kapusta_photons} have found
a $\pi\rho \rightarrow \pi\gamma$ reaction with intermediate virtual pion or
rho
to be the main source of photons with energy greater than 0.7 GeV.
 However, at temperatures
considered T=100-200 MeV, a $\pi\rho$ pair
can easily form an $A_1$(1260) resonance, and we
show that this mechanism
leads to the photon production  rates exceeding those
suggested previously.

\vskip .25cm
\vfil
\eject

\newpage
\pagestyle{plain}
\setlength{\baselineskip}{16pt}
\vfill\eject
\centerline{\bf 1. Introduction}
\medskip
    Theoretical and experimental studies of very dense and hot hadronic matter
is one of the most active fields, recently created at the intercept of nuclear
and high energy physics. One of the goals of this program is to
 find evidence for phase transitions into new form of matter -- the
quark-gluon plasma.

   As suggested in \cite{Shuryak_80}, a promising way to detect these
phenomena experimentally is to look at
electromagnetic radiation of photons and lepton pairs, the `penetrating probes'
which
do not suffer the final state interaction. A number of works
\cite{Kajantie_photons,Halzen_photons,Sinha_photons,Hwa_photons,Staadt_photons}
were devoted during the last decade to evaluation of the corresponding
emission rates from quark-gluon plasma.

  However, expanding hadronic matter spends most of its time
in the form of rather cool hadronic gas, which also
 produces some electromagnetic radiation.
Kapusta et al \cite{Kapusta_photons}
have recently attempted to calculate the rate for photon emission,
using a model Lagrangian describing interaction of $\pi,\rho$ mesons
with photons. It was pointed out in this work, that out of all
possible processes involved, the $\pi\rho \rightarrow \pi\gamma$ reaction
plays the dominant role in yielding photons with energies larger than
0.7 GeV.

    In a different context, one of us \cite{Shuryak_pot}
has recently discussed a number of hadronic reactions in a hadron gas.
It was similarly found, that the main effect of $\rho$ meson modification
in hadronic gas
is its scattering on pions. Moreover, comparing various contributions
to this scattering, it was found that the one related with the $A_1$
resonance is much more important, than that with an intermediate pion.
Since only the latter was included by Kapusta et al, the question was raised
whether the $A_1$-related process can also produce an important (or even
dominant) contribution to the photon production. As will be shown in the
present
work, it is indeed the case and  the $\pi\rho \rightarrow
A_1 \rightarrow \pi\gamma$ reaction ( see Fig.(1) ) does `outshine'
all others.

   The paper is organized as follows. In section 2 we describe the
possible form of $\pi\gamma A_1$ interaction. The corresponding coupling
constant is estimated in the vector dominance model. Then we evaluate the
partial width $\Gamma(A_1\rightarrow\pi\gamma) $, and compare it with other
estimates and experimental data. Another test of these estimates are
provided by the {\it pion polarizability}, for which also some experimental
and theoretical estimates are available in the literature.  We have found
reasonable agreement with the theoretical estimates,
while the experimental
situation is extremely uncertain and even contradictory.
   In section 3 we evaluate photon production rate due to the mechanism
considered, and compare it with results obtained by Kapusta et al.

\bigskip
\centerline{\bf 2. The  $\pi\gamma A_1$ interaction }
\medskip

  The process we are going to discuss is
 $\pi\rho\rightarrow A_1 \rightarrow \pi\gamma$,
see Fig.(1), and its
total cross section in the region of $A_1$ resonance
can be written as
\be
\sigma (\sqrt s) = {\pi\over {\bf p}^2}
{\Gamma_{A_1\rightarrow\pi\rho}\Gamma_{A_1\rightarrow\pi\gamma}
\over (\sqrt s - m_{A_1})^2 + \Gamma_{A_1}^2/4 }
\label{eq:sig1}
\ee
where ${\bf p}$ is the three momentum of the rho in the c-m frame.
$\Gamma_{A_1}$
is the total width of $A_1$ and is approximately equal to
$\Gamma_{A_1\rightarrow\pi\rho}$.

The axial $A_1$ meson is a strong resonance known for about 3 decades,
 and (although its properties \cite{awidth} measured
in hadronic reactions
and $\tau \rightarrow \nu_\tau+A_1$ \cite{Schmidke_taodecay} are slightly
different) its total and partial width
$\Gamma_{A_1\rightarrow\pi\rho}$ are reasonably well known.

  However, the radiative decay width $\Gamma_{A_1\rightarrow\pi\gamma}$ is
far from being firmly determined.
Its value given by Particle Data Tables
is based on one experiment only, and there seems to be some problems with
it (see below). Therefore
we try to estimate its magnitude theoretically, and also compare
the results with as many previous theoretical and experimental
on the subject as possible.

By using the vector dominance model (VDM) \cite{Sakurai}, we can
approximately relate
the radiative decay width to the  width to $\pi\rho$:
\be
\Gamma_{A_1\rightarrow\pi\gamma}=
\Gamma_{A_1\rightarrow\pi\rho}
\label{eq:approx} ({e\over f_\rho})^2 \ee
where $e$ is the electron charge and
$f_\rho$ is the well known coupling constant of $\rho NN$ and $\rho \pi\pi$
(from the $\rho\rightarrow \pi\pi$ width, we find it be
${f_\rho^2 \over 4\pi}= 2.9$).

  Of course, Eq.(\ref{eq:approx}) is a rough estimate, and one may try to
improve it by
taking into account the ratio of the phase space in both reactions.
Also since $A_1$ is a wide resonance,
 the ratio of the matrix elements
 is modified  when the total invariant mass of the
process $\sqrt s$ is shifted away from the centroid of the resonance.
Particularly in our thermal production
problem, the region near the threshold, at
 $\sqrt s < m_{A_1}$, is
especially important,  since it is strongly
 favored by  the thermal distribution.
To evaluate those corrections, one has
to study the $A_1$ physics within some Lagrangian model.

In constructing the $\pi\gamma A_1$ Lagrangian,
 we note that it is a subject for two general constraints:
(i) gauge invariance with respect to the photon field; (ii)
the interaction of soft pions
 should be proportional to either their momenta or quark masses,
since the pion is a Goldstone boson related with chiral symmetry breaking.
Let the fields for $A_1$, photon, and pion be denoted as $a^\mu$, $A^\mu$,
and $\phi$ respectively; then
we can write the Lagrangian as
\be
 L_{\pi\gamma A_1}
&=& G_\gamma  a^\mu \Gamma_{\mu\nu} A^\nu \phi \label{eq:def}\\
&=& G_\gamma a^\mu
( g_{\mu\nu} p_\pi\cdot p_\gamma - p_{\gamma \mu} p_{\pi \nu} ) A^\nu \phi .
\label{eq:lagrangian}\ee
In the above
$p_\pi$ and $p_\gamma$ are four momenta
carried by the pion and the photon.
With the above  two constraints,
Eq.(\ref{eq:lagrangian}) is the
only possible structure of the Lagrangian with minimal number of derivatives.

The effective coupling strength $G_\gamma$ (which bears
the dimension  of inverse mass) can be estimated as above,
 by using the VDM \cite{Sakurai}. This implies that the
$\pi\rho A_1$ Lagrangian has similar form as that of $\pi\gamma A_1$, namely
\be\label{rho}
 L_{\pi\rho A_1}
= G_\rho \rho^\mu
( g_{\mu\nu} p_\pi\cdot p_\rho - p_{\rho \mu} p_{\pi \nu} ) A^\nu \phi
\label{eq:rho}
\ee
where $\rho^\mu$ is the field for the rho and $p_\rho$ is its four-momentum.

The $\pi\rho A_1$ Lagrangian has been studied before in the chiral gauge
model, first by Schwinger \cite{Schwinger}, and further by
others \cite{Zumino,Nieh}.
Note that their Lagrangian looks as
the first term in Eq.(\ref{eq:rho}), but in this case,
when  rho is substituted by gamma, it violates the
gauge invariance and therefore cannot be used for our
 VDM-type estimates. \footnote{The real accurate Lagrangian
can only be made if the quantitative measurements of
 s- and d-wave contributions be made.
To construct the accurate Lagrangian requires the experimental measurement
of the angular distribution of the $A_1$ decay. By choosing $\theta$ to be
the angle between the $A_1$ polarization and the moving rho in the $A_1$
rest frame, our model yields
$$ f(\theta) \sim 1+ 0.946 \cos^2\theta $$   }

The $A_1\rightarrow \pi\rho$ decay width is then found
to be
\be
\Gamma_{A_1\rightarrow\pi\rho}=
 {G_\rho^2 |{\bf p}|\over 24\pi m_{A_1}^2} \left[ 2(p_\pi\cdot p_\rho )^2
 +m_\rho^2 (m_\pi^2+{\bf p}^2) \right]
\ee
where ${\bf p}$ is the pion momentum in the rest frame of $A_1$.
 From the above equation we determine
$ G_\rho = 14.8 \ GeV^{-1}$, using the averaged experimental value for
the total width $\Gamma_{A_1}=0.4$ GeV.

The coupling constants $G_\gamma$ and $G_\rho$ relate to each other in the VDM
simply by
\be  G_\gamma= G_\rho {e\over f_\rho}=0.743\ {\rm GeV}^{-1}. \ee
With this we predict the
radiative decay width
\be\Gamma_{A_1\rightarrow\pi\gamma} =
{G_\gamma^2 |{\bf p}|\over 12\pi m_{A_1}^2}(p_\pi\cdot p_\gamma)^2
=1.42\ {\rm MeV}. \ee

Now we are going to check whether this value produces results consistent
with the experimental data and other theoretical estimates.
The theoretical evaluation of the $A_1$ radiative
width has been carried out in the
framework of the non-relativistic SU(6) quark model in \cite{Rosner_SU6}.
Their result
\be \Gamma_{A_1\rightarrow\pi\gamma} =  1.0 \sim 1.6\ {\rm MeV}. \ee
is quite consistent with our estimate.

The experimental evidence for the radiative decay of $A_1$ was found
through its reverse process, i.e., by studying $A_1$ production in
pion-nucleus collisions.
With two different targets, lead and copper,
the experiments were performed and yield the partial width \cite{Zielinski}
\be
\Gamma_{A_1\rightarrow\pi\gamma}=  0.640\pm 0.246\, {\rm MeV}, \ee
 assuming
the reaction is purely electromagnetic. However, in this case the cross section
of the $A_1$ production should be proportional to $Z^2$,
with $Z$ the charge of
the target. This dependence is not clearly observed in the data, which
implies that strong interaction somewhat  interferes with the electromagnetic
one and can produce some systematic errors.

The pion polarizability sheds some light on
$A_1 \pi\rho $ interaction as well. Theoretically
it was studied through
the low energy Compton scattering on the pion \cite{Holstein,Petrunkin}.
The forward scattering amplitude relates with the electric
and the magnetic polarizabilities $\alpha_E, \beta_M$ through \cite{Friar}
\be f(0)= {1\over 4\pi}({\bf
\epsilon\cdot\epsilon^\prime} w w^\prime \alpha_E
+ {\bf \epsilon\times q \cdot \epsilon^\prime\times q^\prime} \beta_M)
\ee
where $(w,{\bf q}), {\bf \epsilon}$ and $(w^\prime,{\bf q^\prime}),
{\bf \epsilon^\prime}$ are
the four momenta
and polarizations of the incident and the outgoing photon respectively.
In the chiral limit, it has been shown \cite{Holstein} that the pion electric
polarizability from exchanging the vector current cancels the part from the
pion formfactor, and therefore the process of intermediate $A_1$ is believed to
dominate the pion polarizability.

With our estimated parameters, the
 pion electric polarizability from $A_1$ contribution
is found to be
\be\alpha_E = {G_\gamma^2 m_\pi\over 2m_{A_1}^2}
= 1.8\times 10^{-4} \, {\rm fm}^3. \ee
This value should be compared to
the former QCD-based calculation \cite{Holstein_1}
which yields $2.8\times 10^{-4} \ {\rm fm}^3$, and the result of
the current-algebra  calculation \cite{Holstein},
$2.6\times 10^{-4} \ {\rm fm}^3$.

The experiments give two conflicting  values for the
pion electric polarizability. Both are {\it larger} than the theoretical
results:
\be \alpha_E \approx (6.8\pm 1.4)\times 10^{-4} \ {\rm fm}^3 \ee
 from radiative pion scattering \cite{Antipov}
and
\be \alpha_E \approx (20\pm 12)\times 10^{-4} \ {\rm fm}^3 \ee
 from the pion photoproduction \cite{Aibergenov}.
The discrepancy of the theoretical predictions and the experimental
measurements remains unsolved.

\bigskip
\centerline{\bf 3. Photon Production from Hadronic Gas}
\medskip
Photon production from the hot hadronic phase was studied recently
by Kapusta et al \cite{Kapusta_photons}, who
have calculated the photon production from the light ($\pi,\rho$) meson
interactions.
They have found that
the pion-rho collision channel dominates
for photons with energy greater than 0.7 GeV.

 Including $A_1$ as an intermediate  state, one
can generally add many diagrams to photon production processes, but
from kinematic arguments it is clear that only the s channel $\pi\rho$
interaction is significant,
since in the hadronic gas of $T= 100-200$ MeV the total energy of a pion
and a rho combined is right at the $A_1$ resonance.

The photon production rate from a hadronic
gas with  temperature $T$
can then be calculated from
\def\rr #1 #2 {\displaystyle{ d^3#1_#2 \over (2\pi)^3 2E_#2 }}
\be\displaystyle
E_\gamma {dR\over d^3p_\gamma}&=&{\cal N}\int \rr p 1 \rr p 2 f(E_1) f(E_2)
 \sum_{\lambda,\sigma}|{\cal M}^{\lambda,\sigma}|^2 \nonumber\\
&&(2\pi)^4 \delta^4 ( p_1+p_2-p_3 -p_\gamma )
{d^3p_3\over (2\pi)^6 4E_3}   (1+f(E_3)).
\label{eq:rate}
\ee
In the above the indices $1, 2, 3,$ and $\gamma$ are for incident pion,
incident rho,
outgoing pion, and outgoing photon respectively; ${\cal N}$ is the isospin
degeneracy of the process;
The matrix element for the process is
\be {\cal M}^{\lambda,\sigma}= G_\rho G_\gamma
\epsilon_\rho^{\mu,\lambda} \Gamma_{\mu\alpha} D^{\alpha\beta}
\Gamma_{\nu\beta} \epsilon_\gamma^{\nu,\sigma}
\ee
where $\epsilon_\rho, \epsilon_\gamma$ are the polarization of the
incident rho and outgoing photon; $\lambda, \sigma$ are their spin indices;
The vertex structure function
$\Gamma_{\mu\nu}$ is defined by Eq.(\ref{eq:def});
$D^{\alpha\beta}$ is the propagator for $A_1$
\be
 D^{\alpha\beta}(p) = (g^{\alpha\beta}- p^\alpha p^\beta / m_{A_1}^2)
{1\over p^2- m_{A_1}^2 -i m_{A_1} \Gamma_{A_1} };
\ee
$f(E)= 1/( exp(E/T)-1)$ is the Bose-Einstein distribution function.
We replace it by the
Boltzmann distribution, since we are interested in photons with energy
much greater than the temperature.
With this approximation, the right-hand side of Eq.(\ref{eq:rate})
can be simplified as a three-dimensional integral.
We numerically computed the integral and get the photon spectra.
With the matrix elements of
\cite{Kapusta_photons} due to exchanging intermediate pions and rhos,
we have also reproduced their spectra, which are shown in Fig.(2) for
comparison (the dashed curves).
Our results for photon production through
the $A_1$ resonance
are shown in Fig.(2) by solid curves. We have calculated the spectra
for three different temperatures $T=$ 200, 150, and 100 Mev, and
for all curves, the combination of the thermal
distribution and the matrix element results in the photon production
peaking at $0.5\sim 0.7$ GeV.
The pion-rho scattering processes dominate the photon production
 for $E_\gamma > 0.7$ GeV; while  photons with smaller energy
come from other sources.

One can see from Fig.(2),
 that in the pion-rho dominated region, the photon production from
 the $A_1$ resonance is consistently greater than that from
the diagrams with virtual pions and rhos.
This is true even at the lowest temperatures considered, and
  with increasing T the effect is naturally more pronounced.
At the same time, as one approaches the critical temperature of chiral
restoration, the calculation becomes more uncertain because of possible
strong modification of mesons, especially masses of mesons in matter.
It has
been proposed \cite{Brown_mass} that meson masses, other than the mass of the
pion, decrease with temperature, going essentially to zero at the temperature
for chiral restoration $T_c=T_{\chi SR}$. This temperature is now
estimated to  be $T_c\sim 140$ MeV in lattice gauge calculations
\cite{Bernard_lattice}. The number of $\rho$ mesons will be greatly increased
if $m_\rho^*$ drops, since Boltzmann factors will be larger, and this would
strongly increase the process $\pi\rho\rightarrow \pi\gamma$,
while the increase should bias
at low energy photons more than at high energy photons.
If one also takes into account the expected decrease
in the $A_1$ mass, the photon production
at high T  becomes even larger.

Let us also note that two mechanisms considered above, those due to
 virtual $\pi,\rho$ and $A_1$ mesons, have some overlap in terms
of partial waves, so after integration over angles
some  interference
effect should appear. It may result in some small changes in the
photon production rate. We have not calculated such details
because due to  uncertainties involved it does not make much sense.

Recently there are some efforts devoted to evaluation of
 photon production using the realistic space-time picture of the collision,
especially using the hydrodynamic model \cite{Chakrabarty_photons}.
Photons are
accumulated at each temperature stage of the expansion of the fireball
formed after the heavy-ion collisions.
For the convenience of these calculations, we parametrized the
rate of photon production ( Eq.(\ref{eq:rate}) ) by a simple analytic
function. Following \cite{Nadeau_fit} we suggest the parametrization to be
\be
E_\gamma{dR\over d^3p_\gamma} = 2.4*T^{2.15}exp[-1/(1.35TE_\gamma)^{0.77}
-E_\gamma/T ] \ \ ({\rm fm}^{-4}{\rm GeV}^{-2}),
\label{eq:fit}
\ee
here $E_\gamma$ and $T$ should be in units of GeV.
We marked the parametrization by the solid squares in Fig.(2) for
different temperatures. When comparing with the exact values given by
the solid curves, we see that Eq.(\ref{eq:fit}) is an excellent
fit for the production rate.

\bigskip
\centerline{\bf Acknowledgement}
\medskip
We would like to thank P. Lichard for presenting some details
of his calculations.
We also thank M. Prakash for his helpful comments.
This work is supported in part by the US Department
of Energy under Grant No. DE-FG02-88ER40388.


\newpage
\pagestyle{empty}
\begin{center}
\centering{\bf{\large Figure Captions}}
\end{center}
\vspace{1.3cm}
\begin{description}
   \item [{Figure 1:}]
Feynman diagram of $\pi\rho\rightarrow\pi\gamma$ through
$A_1$ resonance.
   \item [{Figure 2:}]
Energy differentiated rate of photon production
from $\pi\rho\rightarrow\pi\gamma$ processes in hot hadronic gas
of temperature T= 200, 150, and 100 MeV (top, middle, and bottom).
The solid curves are contribution from mechanism of exchanging $A_1$;
the solid squares are from the suggested
parametrization which fits the solid curves; the dashed curves are photon
production from exchanging virtual pions and rhos.
\end{description}

 \end{document}